\documentclass[onecolumn, number, 5p]{elsarticle}
\biboptions{merge,elide}
\usepackage[T1]{fontenc}
\usepackage[utf8]{inputenc}
\usepackage[english]{babel}
\usepackage{lmodern, amsmath, amsfonts, cancel, graphicx}
\usepackage[colorlinks=true, citecolor=blue, urlcolor=blue, linkcolor=blue, breaklinks=true]{hyperref}
\usepackage{booktabs,multirow,tabularx}
\usepackage{axodraw4j}
\usepackage{color}
\usepackage{url}
\usepackage{natbib}

\usepackage{graphicx,amsmath,amssymb,slashed}
\usepackage{epsfig}
\usepackage{pstricks,rotating, graphicx}
\usepackage{latexsym}
\usepackage{booktabs}
\usepackage[Q=yes,pverb-linebreak=no]{examplep}

\usepackage{color}
\usepackage{colordvi}
\usepackage[hang]{subfigure}

\DeclareGraphicsExtensions{.eps}
\graphicspath{{./}}

\newcommand{\tev}{\,\textrm{TeV}}

\begin{document}

%%%%%%%%%%%% Begin Cover Page %%%%%%%%%%%%%%%%%%%%%%%%%%%%%%%%%%%%%%%%%%

\author[a]{R.~Frederix}
\author[a]{S.~Frixione} 
\author[b]{V.~Hirschi} 
\author[c]{F.~Maltoni}
\author[c]{O.~Mattelaer}
\author[d]{P.~Torrielli}
\author[c]{E.~Vryonidou}
\author[e,f]{M.~Zaro}

\address[a]{PH Department, TH Unit, CERN, CH-1211 Geneva 23, Switzerland}
\address[b]{SLAC, National Accelerator Laboratory\\
2575 Sand Hill Road, Menlo Park, CA 94025-7090, USA}
\address[c]{Centre for Cosmology, 
 Particle Physics and Phenomenology (CP3),\\
 Universit\'e Catholique de Louvain, B-1348 Louvain-la-Neuve, Belgium}
\address[d]{Institut f\"ur Theoretische Physik, Universit\"at Z\"urich,  CH-8057 Z\"urich, Switzerland}
 \address[e]{Sorbonne Universit\'es, UPMC Univ. Paris 06, UMR 7589, LPTHE, F-75005, Paris, France}
 \address[f]{CNRS, UMR 7589, LPTHE, F-75005, Paris, France}

\title{Higgs pair production at the LHC with NLO and parton-shower effects}

\begin{abstract} 
We present predictions for the SM-Higgs-pair production
channels of relevance at the LHC: gluon-gluon fusion, VBF, and top-pair, $W$,
$Z$ and single-top associated production. All these results are at the NLO
accuracy in QCD, and matched to parton showers by means of the MC@NLO method;
hence, they are fully differential. With the exception of the gluon-gluon
fusion process, for which a special treatment is needed in order to improve
upon the infinite-top-mass limit, our predictions are obtained in a fully
automatic way within the publicly available {\sc MadGraph5\Q{_}aMC@NLO}
framework.  We show that for all channels in general, and for gluon-gluon
fusion and top-pair associated production in particular, NLO corrections
reduce the theoretical uncertainties, and are needed in order to arrive at
reliable predictions for total rates as well as for distributions.  
\end{abstract}

\maketitle

\section{Introduction}
\label{sec:intro}

Present LHC data provide evidence that the scalar particle observed at the LHC
is the one predicted by the Brout-Englert-Higgs symmetry breaking
mechanism~\cite{Englert:1964et,Higgs:1964pj} of $SU(2)_L \times U(1)_Y$ as
implemented in the Standard Model (SM)~\cite{Higgs:1964pj}.  In this case, the
strengths of the Higgs boson couplings are uniquely determined by the masses
of the elementary particles.  The measured couplings to fermions and vector
bosons agree within 10-20\% with the SM
predictions~\cite{cms2013,atlas2013}. No information, however, has been
collected so far on the Higgs self-coupling $\lambda$.  In the SM the Higgs
boson mass itself fixes the value of this self coupling in the scalar
potential whose form, in turn, is determined by the global symmetries and the
requirement of renormalisability.  These conditions, however, have no 
{\it raison d'\^ etre} once experimental indications (such as the existence 
of dark matter) as well as theoretical arguments (such as naturalness) are put
forward.  In this respect, it is appropriate (and, in fact, advantageous) to
consider the SM as the subset of operators of dimension less than or equal to
four of an effective field theory (EFT) lagrangian with an $SU(2)_L \times
U(1)_Y$ symmetry.  Direct information on the Higgs three- and four-point
interactions could therefore provide a key indication of the structure of the
scalar potential, and of where the scale $\Lambda$ characterising such 
an EFT might lie.

In this context, Higgs pair production could play a key role. Not only it is the
simplest production process that is sensitive to the self-coupling $\lambda$,
but it also provides one with a wealth of possibilities for probing
higher-dimensional interactions as well as the existence of heavier states
coupled to the
Higgs~\cite{Asakawa:2010xj,Dawson:2012mk,Contino:2012xk,Kribs:2012kz,
Dolan:2012ac,Dolan:2012rv,Gouzevitch:2013qca}. Unfortunately, in the SM the
rates for Higgs pair production at the LHC are quite
small~\cite{Plehn:1996wb,Dawson:1998py,Binoth:2006ym,Baglio:2012np}. 
So unless new physics
produces sizable enhancements (something quite possible in several scenarios),
a measurement of the $HH$ production cross sections will necessitate 
considerable integrated luminosity even at 14 TeV centre-of-mass energy. In
any case, precise predictions for rates and distributions will be needed 
in order to be able to extract valuable information on $\lambda$ or on new
physics effects in general.

Analogously to single-Higgs production, several channels can lead to a final
state involving two Higgs bosons. They entail the Higgs coupling to either the
top quark (as in the case of gluon-gluon fusion and of $t \bar t$ associated
production), or vector bosons (in VBF, and in $W$ and $Z$ associated
production), or both (for single-top associated production).  The dominant
production mechanism is gluon-gluon fusion via a top loop, exactly as in the
case of single-Higgs production.  Cross sections corresponding to the other
channels are at least one order of magnitude smaller, even though possibly
interesting because of different sensitivity to $\lambda$ or to new physics,
and because of the possibility of exploiting a wider range of Higgs decay
signatures.

In this Letter we present results accurate to NLO in QCD for the six 
production channels mentioned before, which are the largest in the SM. 
For {\em all} of them our predictions improve upon existing ones in at 
least one aspect. We shall discuss this point in more details in what 
follows. Here, we limit ourselves to pointing out that $HH$ production 
via gluon-gluon fusion is computed at the NLO in a ``loop-improved'' EFT 
approach, using the exact one-loop real-emission and improved one-loop virtual
matrix elements; that in the case of $t\bar{t}HH$ and $tjHH$ production
exact NLO QCD results are presented in this paper for the first time; and
that by matching NLO computations to parton showers we generate samples
of events, also for the first time, for each of the production channels, 
which can be used for fully realistic simulations, including those at detector
level.  With the exception of the gluon-gluon fusion process which, being
loop-induced, needs an ad-hoc treatment, our results are obtained
automatically with the publicly-available version of the {\sc
MadGraph5\Q{_}aMC@NLO} framework~\cite{MG5_aMCpap,MG5_aMC}.

In the next section we introduce and review the main features of the 
Higgs-pair production channels. In section~\ref{sec:method} we present the
calculation and simulation framework, and in section~\ref{sec:results} we
collect results for some selected observables together with their 
uncertainties. We summarise our findings and prospects in the conclusions.

\begin{figure}
\begin{center}
\SetScale{0.45}\setlength{\unitlength}{0.45pt}
\fcolorbox{white}{white}{
  \begin{picture}(491,152) (110,-139)
    \SetWidth{1.0}
    \SetColor{Black}
    \Line[arrow,arrowpos=0.5,arrowlength=3.409,arrowwidth=1.364,arrowinset=0.2,double,sep=4](112,-8)(192,-40)
    \Line[arrow,arrowpos=0.5,arrowlength=3.409,arrowwidth=1.364,arrowinset=0.2,double,sep=4](112,-136)(192,-104)
    \GOval(192,-72)(48,16)(0){0.882}
    \SetWidth{1.5}
    \Line[dash,dashsize=10](208,-40)(288,-24)
    \Line[dash,dashsize=10](208,-104)(288,-120)
    \SetWidth{1.0}
    \Line[arrow,arrowpos=0.5,arrowlength=3.409,arrowwidth=1.364,arrowinset=0.2,double,sep=4](368,-136)(448,-104)
    \Line[arrow,arrowpos=0.5,arrowlength=3.409,arrowwidth=1.364,arrowinset=0.2,double,sep=4](368,-8)(448,-40)
    \SetWidth{1.5}
    \Line[dash,dashsize=10](534,-72)(598,-24)
    \Line[dash,dashsize=10](534,-72)(598,-120)
    \Line[dash,dashsize=10](464,-72)(534,-72)
    \BCirc(534,-72){4}
    \SetWidth{1.0}
    \GOval(448,-72)(48,16)(1){0.882}
    \Text(304,-26)[lb]{{\Black{$H$}}}
    \Text(304,-126)[lb]{{\Black{$H$}}}
    \Text(608,-26)[lb]{{\Black{$H$}}}
    \Text(608,-126)[lb]{{\Black{$H$}}}
  \end{picture}
}
 \caption{\label{fig:hh} Classes of diagrams for Higgs pair production in
 hadron hadron collisions: double Higgs production without $HHH$ vertices on
 the left-hand side, and, on the right-hand side, the contribution due to the
 Higgs self interaction.  Final state particles other than the Higgs bosons
 are understood.  }
 \end{center}
\end{figure}
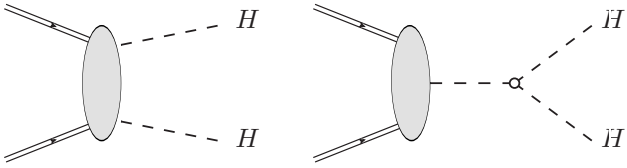

\section{Higgs pair production channels}
\label{sec:channels}

In the SM, the diagrams contributing to Higgs pair production can be organised
in two classes (see fig.~\ref{fig:hh}): those where both Higgs bosons couple 
only to vector bosons or to heavy quarks, and those that feature the Higgs self
coupling.  

The dominant channel for Higgs pair production is gluon-gluon fusion via
virtual top quarks, {\it i.e.}, box and triangle diagrams. This process
therefore starts at the leading order with a loop, exactly as single-Higgs
production.  In contrast with the latter, however, the effective field theory
approach (where Higgs-gluons vertices are included in the lagrangian ${\cal
L}_{\rm HEFT} = \alpha_S/(3\pi v^2) (\phi^\dagger \phi) G G $, $G$ being the
QCD field tensor) provides only a rough approximation for total rates, and a
very poor one for distributions~\cite{Baur:2002rb,Dolan:2012rv}.  
Better predictions, which take loop effects
into account exactly, need therefore to be employed in actual phenomenological 
and experimental studies.  Results have been available for some time, and
implemented in the code {\sc HPAIR}~\cite{Plehn:1996wb,Dawson:1998py}, which
deals with both the SM and the MSSM, but is only capable of computing 
total cross sections.
In {\sc HPAIR} the NLO calculation is essentially performed with EFT 
techniques; the exact one-loop Born amplitudes are however employed
as leading-order contribution to the NLO cross section, and used to
reweight (after the integration over the polar scattering angle)
the HEFT virtual- and real-emission matrix elements. 
In this work we improve on the {\sc HPAIR} approach on several
counts. Firstly, we include the exact one-loop results not only for the
$2\to 2$ Born amplitudes, but also for the $2\to 3$ real-emission processes,
which we compute with {\sc MadLoop}~\cite{Hirschi:2011pa}.
In other words, the only approximation made at the level of matrix elements
is that for the finite part of the two-loop virtual corrections which, being 
presently unknown, is approximated by the corresponding one-loop HEFT 
result reweighted (without any intermediate integration)
with the exact one-loop Born amplitude. 
Secondly, in the loops we make use of the complex mass (and Yukawa) scheme for 
the top quark~\cite{Denner:1999gp,Denner:2005fg}. Thirdly, our results
are fully differential, and can be used to obtain any distribution after matching with parton shower.
In summary, our predictions improve both on the total cross sections that 
can be obtained with {\sc HPAIR}, and on the differential, hadron-level
({\it i.e.}, showered) observables recently presented in ref.~\cite{Li:2013flc,
Maierhofer:2013sha} (which do not include virtual effects, and are 
therefore akin to tree-level merged results). We also stress we do not
make use of the recently-derived $1/m_t$ effects at the NLO 
accuracy~\cite{Grigo:2013rya}, of the NNLO HEFT results for total 
rates~\cite{deFlorian:2013jea} and of threshold resummation~\cite{Shao:2013bz}. More 
details on the procedure 
employed in this work will be presented  elsewhere~\cite{Vetal}.

The second-largest production channel is vector boson fusion (VBF).  In
this case the NLO QCD corrections are trivial, as they involve the same
contributions as for single-Higgs production. In VBF we compute only 
vertex loop corrections, {\it i.e.}, the finite part of the pentagon and hexagon loop
diagrams are discarded for simplicity. These contributions only affect
interferences between diagrams that feature identical quarks, which are
negligibly small already at the LO. NLO results have been presented in the
literature (see e.g.~\cite{Baglio:2012np}) only for total rates. In this 
paper we study, for the 
first time, differential observables for VBF in the SM at fixed NLO 
and matched to parton showers, showing 
distributions for the latter. Distributions at fixed NLO in the two Higgs doublet 
model have appeared in ref.~\cite{Figy:2008zd}. We point out that, although NNLO 
corrections to the total VBF cross sections are not known, they could 
be easily computed following the approach of ref.~\cite{Bolzoni:2010xr}.

At variance with single-Higgs production, the production of a Higgs pair 
in association with a $t\bar t$ pair is the third most important process and, 
in fact, it is even larger than VBF at high Higgs-pair transverse momenta,
or for collider centre-of-mass energies higher than that of the LHC.
The inclusion of NLO QCD corrections in this process has never been achieved
prior to this work, even at a fully inclusive level, as it involves thousands 
of Feynman diagrams of high complexity, such as pentagon and hexagon
loops. Our framework, however, has no problems in handling it in a fully
automatic way. For instance, the total (sequential) CPU time required 
to generate one million unweighted events and to obtain a cross section 
accurate at the per-mil level is about one hundred and sixty CPU hours on a 2.3GHz 
machine. This renders the 
computation feasible on a medium-size (30 core) cluster in a few hours.

The channels of vector boson associated production are technically the 
easiest ones, as all QCD corrections factorise and are relevant only to 
the initial state. As in the case of VBF, we improve upon existing NLO
results by giving one the possibility of studying fully-differential
observables; in this work, we do not include the finite one-loop,
$gg$-initiated contributions to $ZHH$ production, which however can
also be handled by {\sc MadLoop}. Our results correspond to on-shell 
final state vector bosons. NNLO QCD corrections to total cross sections 
are known to be small~\cite{Baglio:2012np}.

Finally, in order to provide the complete set of possibly interesting final
states, we also compute for the first time at the NLO the single-top associated
production, by including both $s$- and $t$-channel contributions and by
considering both top and anti-top in the final state.
The corresponding cross sections are tiny at the LHC, and of
very limited phenomenological relevance in the SM. However, this process is at
least of academic interest because it is sensitive to couplings to both vector
bosons and top quarks, and to their relative phases. In addition to that, 
given that it has the largest sensitivity to the self-coupling $\lambda$, it
might become relevant at a future proton-proton 100 TeV collider.

\begin{figure*}[t]
 \center 
 \includegraphics[width=0.8\textwidth]{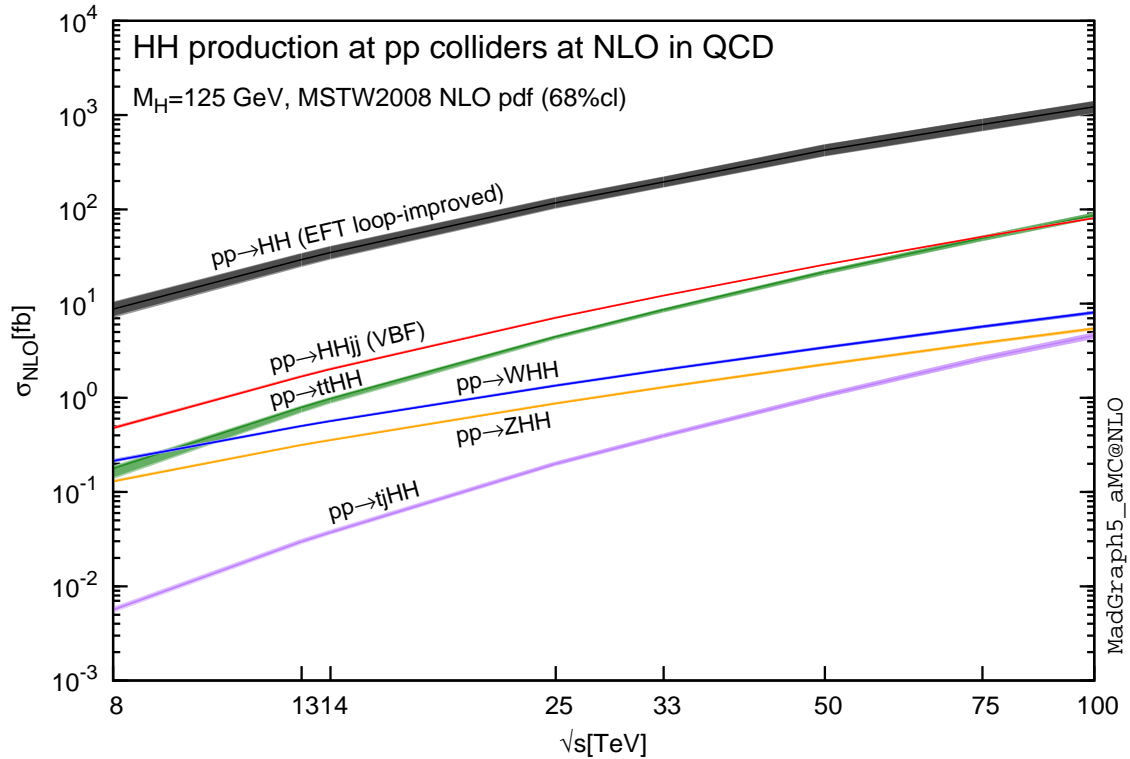}
\vskip.8cm
 \caption{\label{fig:xsec} Total cross sections at the NLO in QCD for the
 six largest $HH$ production channels at $pp$ colliders. The thickness of 
 the lines corresponds to the scale and PDF uncertainties added linearly.}
\end{figure*}

\section{Setup}
\label{sec:method}

As was mentioned above, apart from the gluon-gluon fusion channel, all results
presented in this work have been obtained in a fully automatic way with 
{\sc MadGraph5\Q{_}aMC\-@NLO}~\cite{MG5_aMCpap,MG5_aMC}.  
This program is designed to perform the computation of tree-level and NLO
cross sections, including their matching to parton showers and the merging of
samples with different parton multiplicities.
A user can generate a given process through a simple shell interface 
(in a manner fully analogous to that of {\sc MadGraph5}~\cite{Alwall:2011uj}),
with the corresponding self-contained code being generated on the fly.  
While it is possible to obtain predictions at the ME+PS level ({\it i.e.}, 
with the MLM-$k_T$ tree-level merging technique of 
refs.~\cite{Mangano:2001xp,Alwall:2007fs,Alwall:2008qv}
and its analogues) in this work we limit ourselves to NLO+PS results.
This is because the smallness of the Higgs-pair cross sections rather
emphasises observables which are inclusive with respect to extra radiation,
and for which NLO-level results have to be preferred to tree-level merged
ones, since they provide one with better predictions for absolute 
normalisations and for theoretical uncertainties.

\begin{table*}[t]
\renewcommand{\arraystretch}{1.3}
\begin{center}
    \begin{tabular}{l|cc|cc|cc}
    \toprule
% the header
        \hline
         & \multicolumn{2}{c|}{$\sqrt s = 8 \tev$}
         & \multicolumn{2}{c|}{$\sqrt s = 13 \tev$}
         & \multicolumn{2}{c}{$\sqrt s = 14 \tev$}\\
         & \multicolumn{2}{c|}{ (LO) NLO} 
         & \multicolumn{2}{c|}{ (LO) NLO} 
         & \multicolumn{2}{c}{ (LO) NLO}               \\
         \hline
         $ HH$ {\small (EFT loop-improv.)} & 
$(5.44 ^{+38\%}_{-26\%})  $&$ 8.73 ^{+17 +2.9\%}_{-16 -3.7\%} $ & 
$(19.1^{+33\%}_{-23\%})   $&$   29.3 ^{+15 +2.1\%}_{-14 -2.5\%} $ & 
$(22.8 ^{+32\%}_{-23\%}) $&$   34.8^{+15 +2.0\%}_{-14 -2.5\%} $\\ 
         $ HHjj$ (VBF)  & 
$ (0.436 ^{+12 \%}_{-10 \%}) $&$ 0.479 ^{+1.8 +2.8\%}_{-1.8 -2.0\%} $ & 
$ (1.543 ^{+9.4 \%}_{-8.0 \%})  $&$ 1.684 ^{+1.4 +2.6\%}_{-0.9 -1.9\%} $ & 
$ (1.839 ^{+8.9 \%}_{-7.7 \%})  $&$  2.017 ^{+1.3 +2.5\%}_{-1.0 -1.9\%} $\\ 
$ t\bar t HH $ & 
$ (0.265 ^{+41 \%}_{-27 \%}) $&$ 0.177 ^{+4.7 +3.2\%}_{-19 -3.3\%} $ & 
$ (1.027 ^{+37 \%}_{-25 \%}) $&$ 0.792 ^{+2.8 +2.4\%}_{-10 -2.9\%} $ & 
$ (1.245 ^{+36 \%}_{-25 \%}) $&$ 0.981 ^{+2.3 +2.3\%}_{-9.0 -2.8\%} $\\ 
$ W^+ HH $ & 
$ (0.111 ^{+4.0\%}_{-3.9 \%}) $&$ 0.145 ^{+2.1 +2.5\%}_{-1.9 -1.9\%} $ & 
$ (0.252 ^{+1.4 \%}_{-1.7\%}) $&$ 0.326 ^{+1.7 +2.1\%}_{-1.2 -1.6\%} $ & 
$ (0.283 ^{+1.1 \%}_{-1.3 \%}) $&$ 0.364 ^{+1.7 +2.1\%}_{-1.1 -1.6\%} $\\ 
$ W^-HH $ & 
$ (0.051 ^{+4.2 \%}_{-4.0 \%}) $&$ 0.069 ^{+2.1 +2.6\%}_{-1.9 -2.2\%} $ & 
$ (0.133 ^{+1.5 \%}_{-1.7 \%}) $&$  0.176 ^{+1.6 +2.2\%}_{-1.2 -2.0\%} $ & 
$ (0.152 ^{+1.1 \%}_{-1.4 \%}) $&$  0.201 ^{+1.7 +2.2\%}_{-1.1 -1.8\%} $\\ 
$ ZHH $ & 
$ (0.098 ^{+4.2 \%}_{-4.0 \%}) $&$ 0.130 ^{+2.1 +2.2\%}_{-1.9 -1.9\%} $ & 
$ (0.240 ^{+1.4 \%}_{-1.7 \%}) $&$ 0.315 ^{+1.7 +2.0\%}_{-1.1 -1.6\%} $ & 
$ (0.273 ^{+1.1 \%}_{-1.3 \%}) $&$ 0.356 ^{+1.7 +1.9\%}_{-1.2 -1.5\%} $\\ 
$ tjHH (\cdot 10^{-3}) $ & 
$ (5.057 ^{+2.0 \%}_{-3.2 \%}) $&$ 5.606 ^{+4.4 +3.9\%}_{-2.3 -4.2\%} $ & 
$ (23.20 ^{+0.0 \%}_{-0.8 \%}) $&$ 29.77 ^{+4.8 +2.8\%}_{-2.8 -3.2\%} $ & 
$ (28.79 ^{+0.0 \%}_{-1.2 \%}) $&$ 37.27 ^{+4.7 +2.6\%}_{-2.7 -3.0\%} $\\ 
\hline
\end{tabular}

    \caption{\label{tab:one} LO and NLO total cross sections (in fb) for the
        six largest production channels at the LHC, with $\sqrt{s} = 8, 13, 14
        \tev$. The first uncertainty quoted refers to scale variations, while
        the second (only at the NLO) to PDFs. Uncertainties are in
        percent. No cuts are applied to final state particles and no
        branching ratios are included.}  
\end{center} 
\end{table*}

Within {\sc MadGraph5\Q{_}aMC@NLO}, any NLO computation is performed by
means of two independent 
modules: {\sc MadFKS}~\cite{Frederix:2009yq} takes care of the Born 
and of the real-emission amplitudes, and it also carries out the subtraction 
of the infrared singularities according to the FKS 
prescription~\cite{Frixione:1995ms,Frixione:1997np} as well as the generation
of the Monte Carlo subtraction terms required by the MC@NLO method
\cite{Frixione:2002ik}.  {\sc MadLoop}~\cite{Hirschi:2011pa} computes one-loop
amplitudes, using the OPP integrand-reduction method~\cite{Ossola:2006us} (as
implemented in {\sc CutTools}~\cite{Ossola:2007ax}) and the OpenLoops
method~\cite{Cascioli:2011va}. In the case of VBF and $tjHH$ production, 
some minimal internal manipulations make use of 
{\sc FJcore}~\cite{Cacciari:2011ma}.

In our simulations we set the Higgs mass equal to $m_H=125$~GeV. Parton
distributions functions (PDFs) are evaluated by using the MSTW2008 (LO and
NLO) set in the five-flavour scheme~\cite{Martin:2009iq}.
$b$-quark masses as well as their coupling to the Higgs are neglected.  For
the sake of brevity, we only show observables related to the Higgs bosons and
therefore we have left the latter stable in the simulations.  
We stress, however,
that the top quarks and the vector bosons that appear in the final states can 
be decayed with the built-in {\sc MadSpin} package~\cite{Artoisenet:2012st},
which allows one to include all spin-correlation effects. On the other
hand, Higgs decays can be handled correctly also by the Monte Carlos, 
thanks to the Higgs being a spin-0 particle.

The code allows full flexibility as far as the choice of the renormalisation
and factorisation scales $\mu_{R,F} $ is concerned. The central values
of these scales have been chosen as follows. For gluon-gluon fusion,
VBF, and $VHH$ production we set $\mu_0=m_{HH}/2$, $m_W$ and $m_{VHH}$,
respectively.  For $t\bar t HH$ we choose $\mu_0 = \left(m_T(H_1)\; m_T(H_2)\;
m_T(t) \; m_T(\bar t)\right)^{1/4}$, $m_T$ being the
transverse energy of the corresponding particle, as we find that in this
way the cross section displays a rather stable behaviour.  For single-top
associated production $tjHH$ we simply use the fixed value $\mu_0=m_H+m_t/2$.

Scale and PDF uncertainties can be evaluated at no extra computational cost 
thanks to the reweighting technique introduced in ref.~\cite{Frederix:2011ss}, 
the user deciding the range of variation. In addition, such information is
available on an event-by-event basis and therefore uncertainty bands can be
plotted for any differential observable of interest.  In our analysis we 
vary independently the scales in the range $1/2 \mu_0 <\mu_R,\mu_F <2\mu_0$.  
PDF uncertainties at the 68\% C.L. are obtained by following the prescription 
given by the MSTW collaboration~\cite{Martin:2009iq}.

For the studies shown in this paper we employ 
{\sc HERWIG6}~\cite{Corcella:2000bw} and 
{\sc Pythia8}~\cite{Sjostrand:2007gs} for parton shower and hadronisation.
The matching to {\sc HERWIG++}~\cite{Bahr:2008pv} and 
{\sc Pythia6}~\cite{Sjostrand:2006za} (virtuality ordered, plus $p_T$
ordering for processes with no final-state radiation) 
is also available in {\sc MadGraph5\Q{_} aMC@NLO}.

\begin{figure*}[t]
 \center 
 \includegraphics[width=0.8\textwidth]{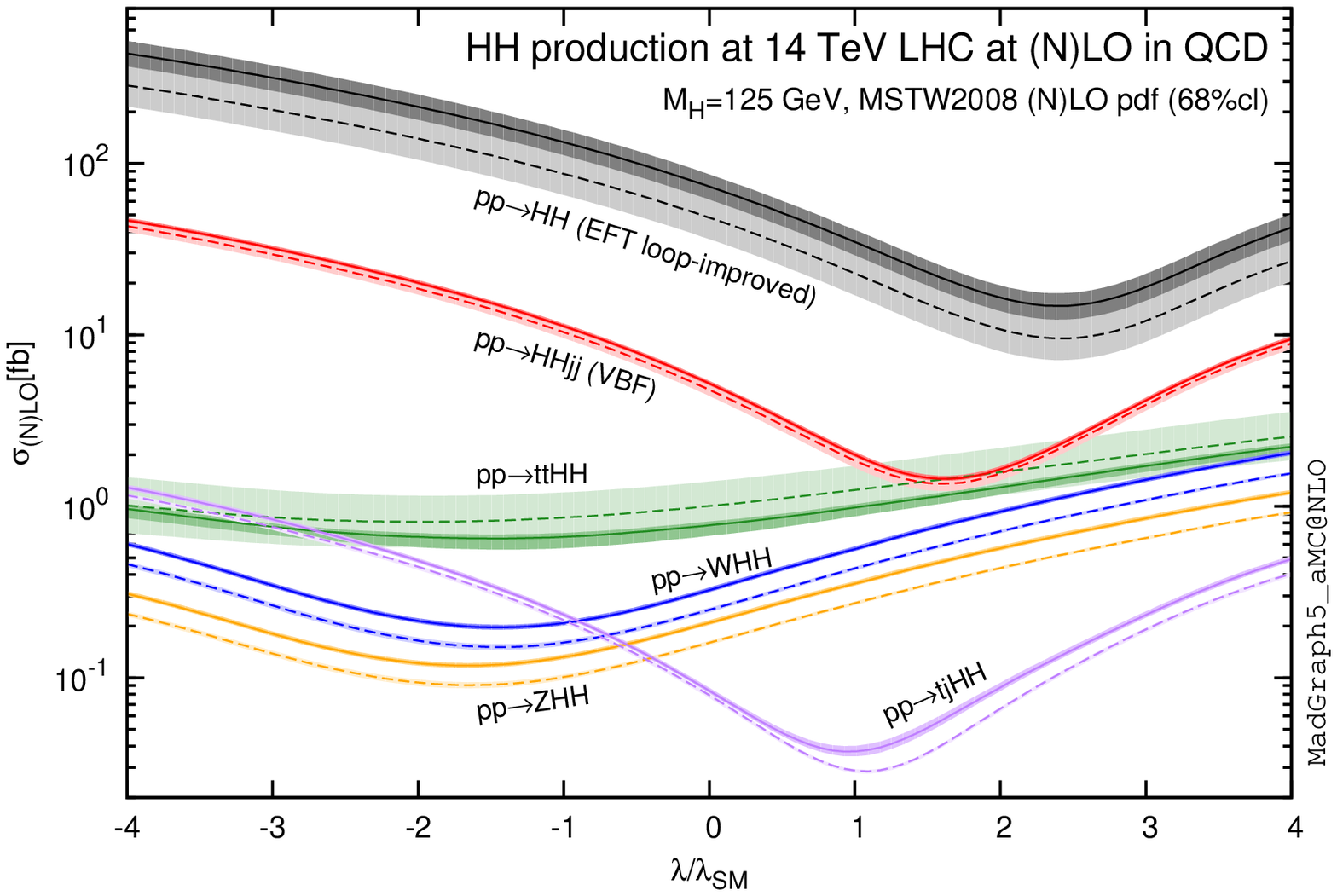}
\vskip.8cm
\caption{\label{fig:xseclam} Total cross sections at the LO and NLO in QCD 
 for $HH$ production channels, at the $\sqrt{s}=$14 TeV LHC as a function 
 of the self-interaction coupling $\lambda$.  The dashed (solid) lines and 
 light- (dark-)colour bands correspond to the LO (NLO) results and to the 
 scale and PDF uncertainties added linearly. The SM values of the cross 
 sections are obtained at $\lambda/\lambda_{\rm SM}=1$.}
\end{figure*}

\section{Results}
\label{sec:results}

We start by presenting in fig.~\ref{fig:xsec} the predictions for the 
total rates at proton-proton colliders with up to 100 TeV c.m.~energy. The
thickness of the curves corresponds to the scale and PDF uncertainties added
linearly. More details are available in table~\ref{tab:one} for selected LHC
energies, {\it i.e.}, 8, 13 and 14 TeV. The first uncertainties (in percent) 
corresponds to scale variation, while the second (only shown at the NLO) to 
PDFs systematics.  Several observations are in order.
Firstly, contrary to what happens in single-Higgs production, the top-pair
associated channel is the third-largest starting at about $\sqrt{s}=$10 TeV,
and becomes the second-largest when c.m.~energies approach $\sqrt{s}=$100 TeV.
Secondly, the theoretical uncertainties due to scale variations in the
three most important processes (gluon-gluon fusion, VBF, and $t\bar{t}$
associated production) are sizably reduced by the inclusion of the NLO
corrections.  Thirdly, the $K$-factor is always slightly larger than one,
except for gluon-gluon fusion where it is of order two, and for the top-pair
associated channel where it is smaller than one. Finally, PDF uncertainties
are comparable to NLO scale uncertainties, except in the case of gluon-gluon
fusion, where the latter are dominant. In the case of $VHH$ and $tjHH$
production it is manifest that the standard procedure of determining
uncertainties due to missing higher orders by varying the scales 
does not give a reliable estimate,
as NLO corrections for these processes are much larger than the LO scale 
dependence band. This is due to two facts: these processes are purely 
electro-weak processes at the LO, and therefore the scale uncertainties
are artificially small; furthermore in the kinematic region probed by these 
processes, the quark-gluon initiated channel which opens up at the NLO can be important.

In fig.~\ref{fig:xseclam} we display total LO and NLO cross sections for
the six dominant $HH$ production channels at the LHC with $\sqrt{s}=$14 TeV, 
as a function of the self-interaction coupling $\lambda$.  The dashed (solid)
lines and light- (dark-)colour bands correspond to the LO (NLO) results and to
the scale and PDF uncertainties added linearly.  The SM value of the cross
section corresponds to $\lambda/\lambda_{\rm SM}=1$. The sensitivity of the
total cross sections to the actual value of $\lambda$ depends in a non-trivial
way on the relative couplings of the Higgs to vector bosons and top quarks,
and on the kinematics in a way that is a difficult to predict a priori, 
{\it i.e.}, without an explicit calculation.  The reduction of the scale
uncertainties that affect the $gg\to HH$, VBF, and $t\bar tHH$ rates, due to
the inclusion of NLO corrections, and pointed out in table~\ref{tab:one} 
for the SM, is seen here also for values of $\lambda\ne\lambda_{\rm SM}$.

\begin{figure*}
 \center 
 \includegraphics[width=0.8\textwidth]{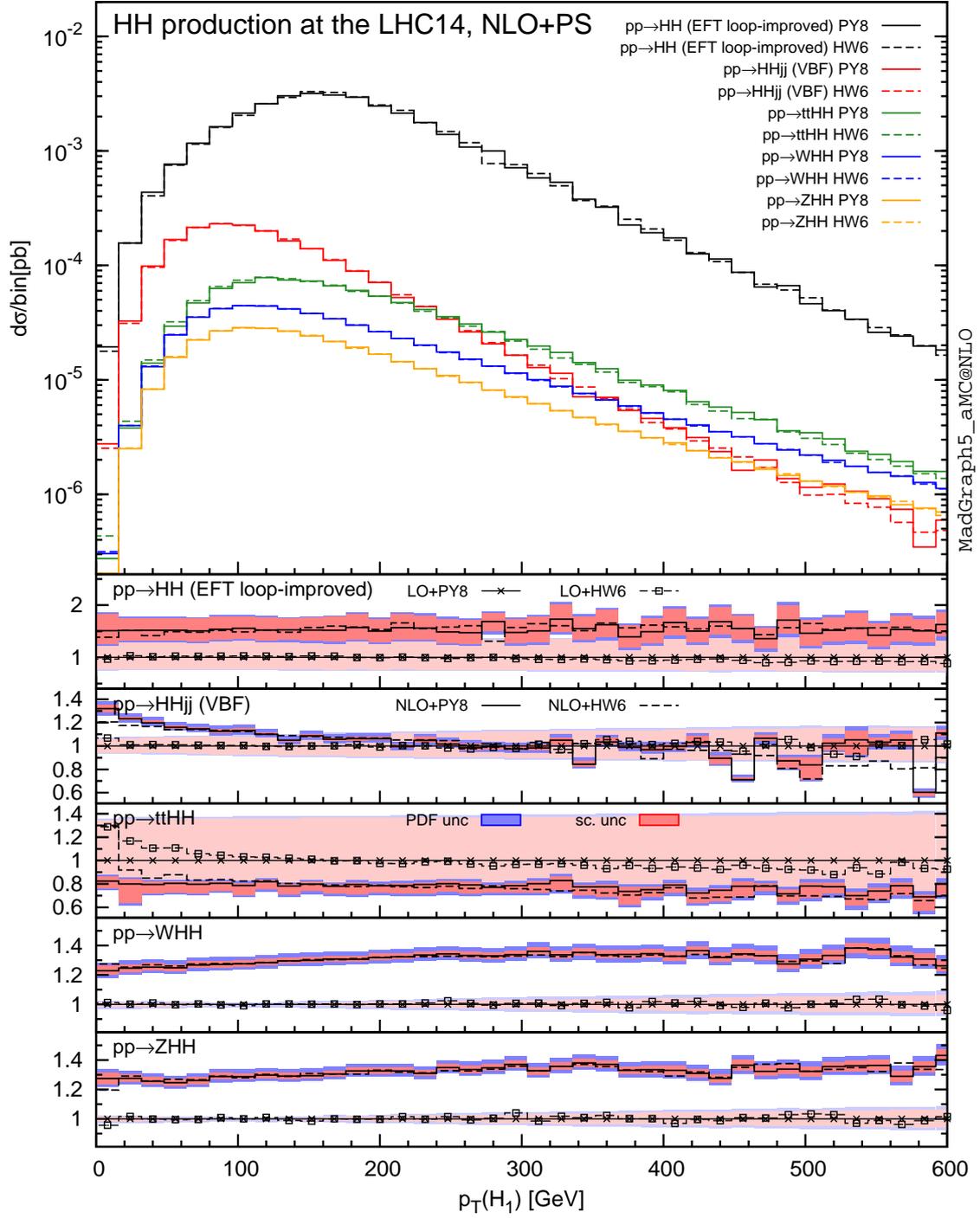}
 \caption{\label{fig:pth1} Transverse momentum distribution of the hardest
 Higgs boson in $HH$ production in the gluon-gluon fusion, VBF, $t \bar t HH$,
 $WHH$ and $ZHH$ channels,  at the 14~TeV LHC. 
 The main frame displays the NLO+PS results obtained after showering 
 with {\sc Pythia8} (solid) and {\sc HERWIG6} (dashes). The insets show, 
 channel by channel, the
 ratios of the NLO+{\sc Pythia8} (solid), NLO+{\sc HERWIG6} (dashes), and
 LO+{\sc HERWIG6} (open boxes) results over the LO+Pythia8 results (crosses).  
 The dark-colour (light-colour) bands represent the 
 scale (red) and PDF (blue) uncertainties added linearly 
 for the NLO (LO) simulations.  }
\end{figure*}

\begin{figure*}
 \center 
 \includegraphics[width=0.8\textwidth]{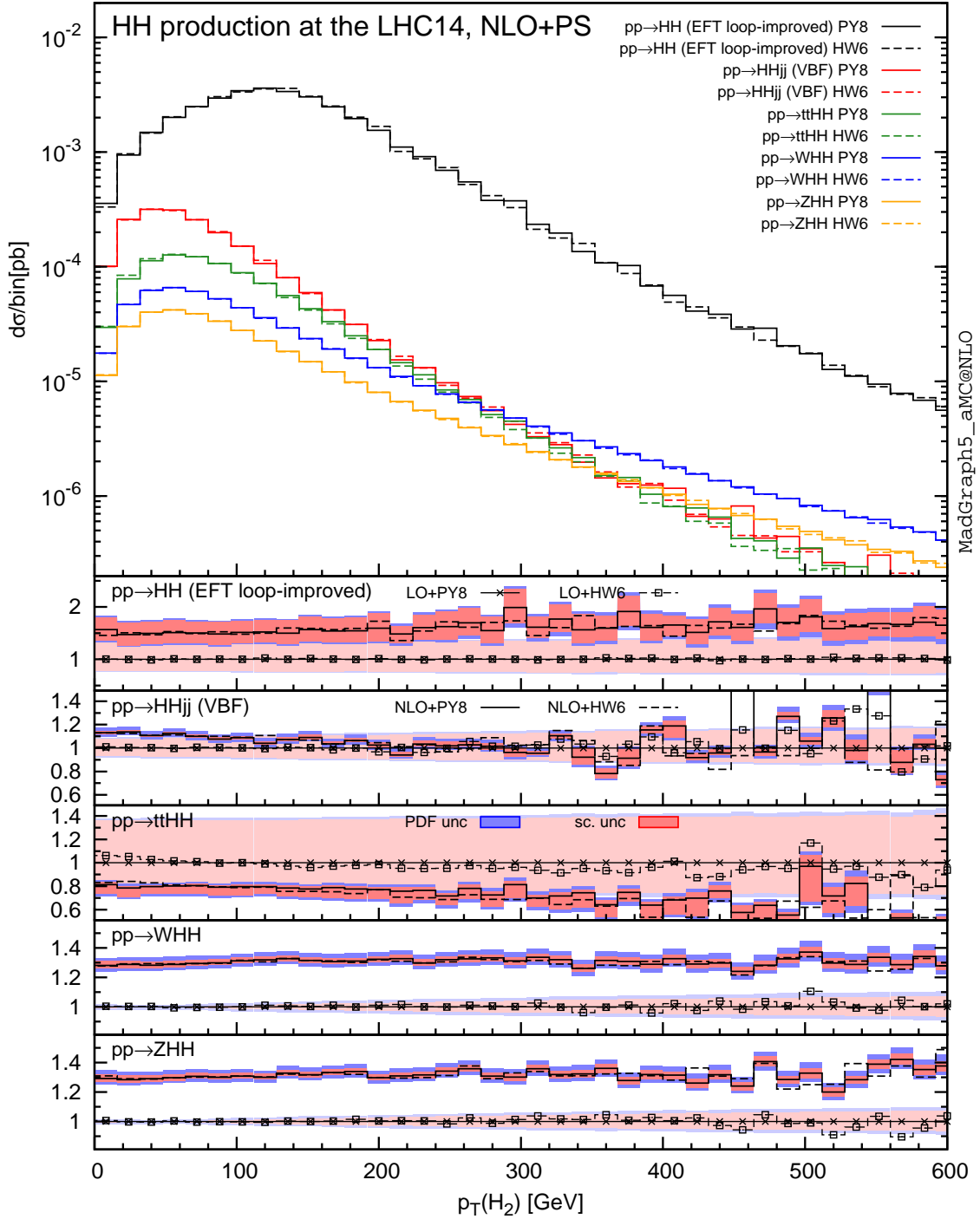}
 \caption{\label{fig:pth2}
 As in fig.~\ref{fig:pth1}, for the softest Higgs bosons. }
\end{figure*}

\begin{figure*}
 \center 
 \includegraphics[width=0.8\textwidth]{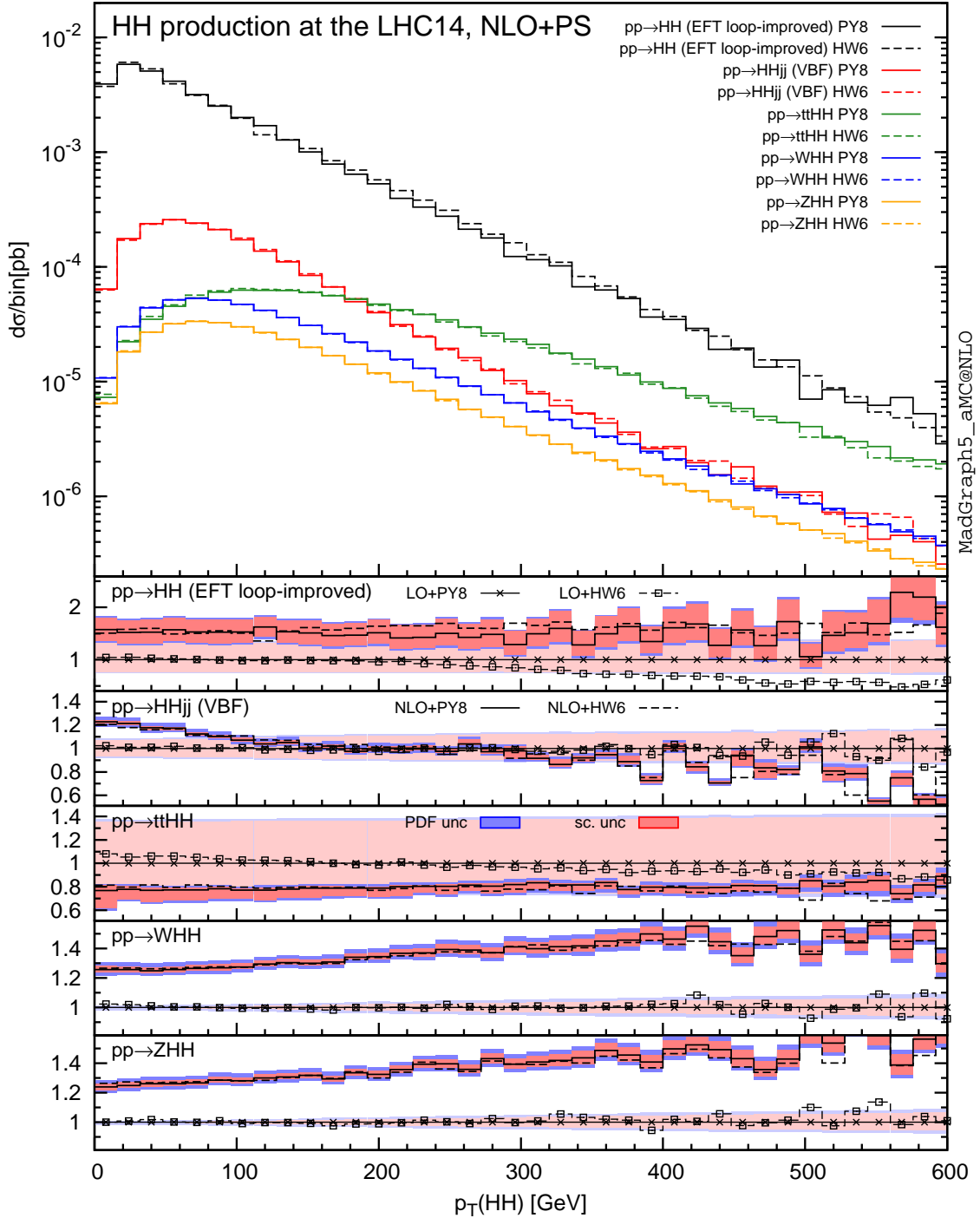}
 \caption{\label{fig:pthh}
 As in fig.~\ref{fig:pth1}, for the transverse momentum of the Higgs pair. }
\end{figure*}

\begin{figure*}
 \center 
 \includegraphics[width=0.8\textwidth]{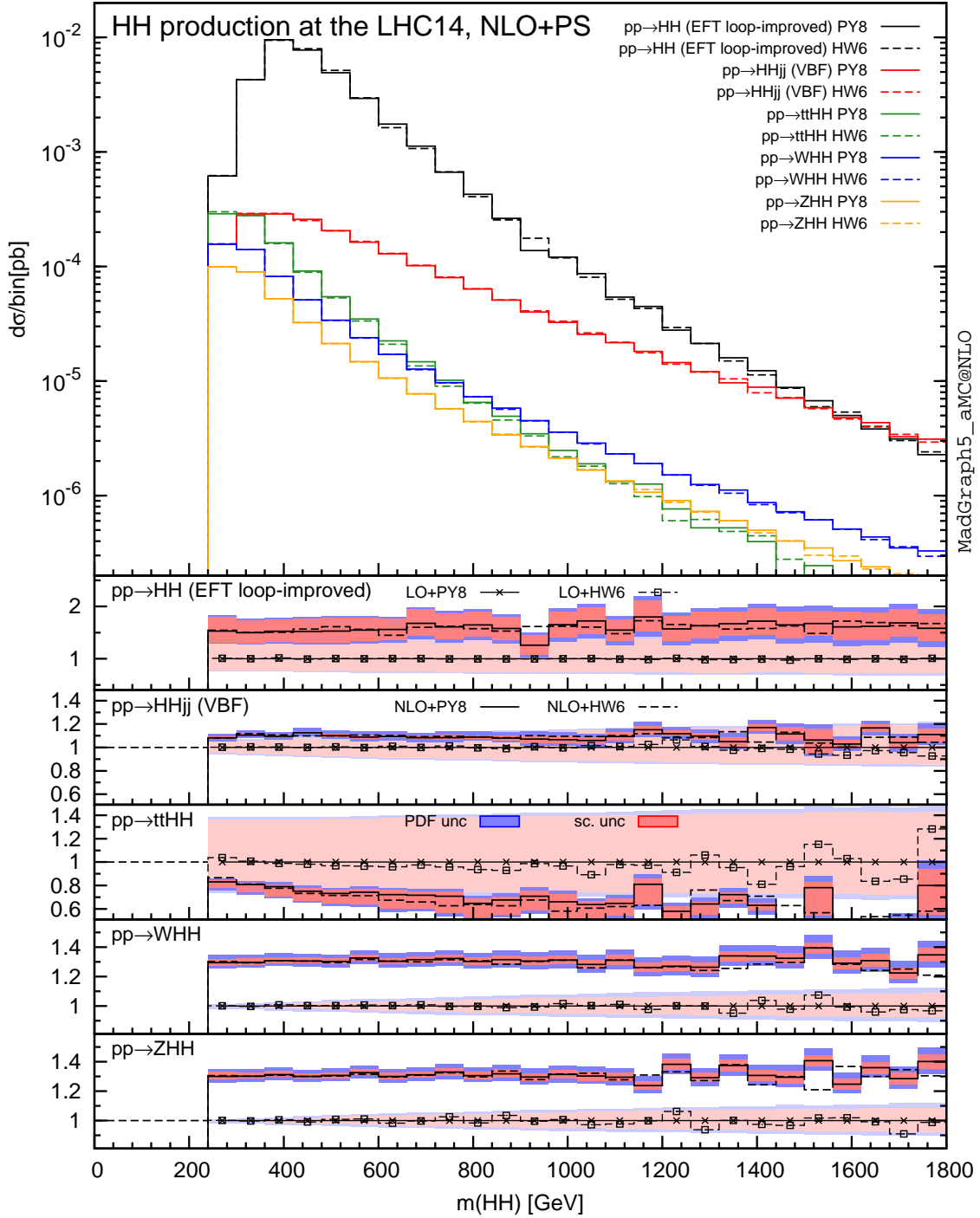}
 \caption{\label{fig:mhh}
 As in fig.~\ref{fig:pth1}, for the invariant mass of the Higgs pair. }
\end{figure*}

We then plot typical distributions for all channels and at the 14~TeV LHC, 
which we obtain by generating samples of  events at parton
level, which are then showered with {\sc Pythia8} (solid) and {\sc HERWIG6}
(dashes).  Being tiny at the 14 TeV LHC, we do not show the results for
single-top associated production. We present observables at the NLO+PS 
accuracy in the main frames of the plots: the transverse momentum of 
the hardest (softest) Higgs in fig.~\ref{fig:pth1} (fig.~\ref{fig:pth2}), and
the transverse momentum (fig.~\ref{fig:pthh}) and the invariant mass 
(fig.~\ref{fig:mhh}) of the Higgs
pair. The insets show, channel by channel, the ratios of NLO+{\sc Pythia8}
(solid), NLO+{\sc HERWIG6} (dashes), LO+{\sc HERWIG6} (dashed with
open boxes) results over the LO+{\sc Pythia8} ones.  The dark-colour 
(light-colour) bands display the scale (red) and PDF (blue) 
uncertainties added linearly for the NLO (LO) simulations.

NLO effects appear as overall rescaling factors only in some distributions and
on a channel-dependent basis. Moreover, differences between results obtained
with the two different shower programs are very mild for all observables and
anyway decreasing when going from LO to NLO. In addition, we have checked that
differences in the distributions between NLO+{\sc Pythia8}/NLO+{\sc HERWIG6}
and NLO fixed-order results are quite small (typically less than a few 
percent), and that in general the NLO+PS results are slightly closer to each
other than to the corresponding NLO fixed-order results. The only exception 
to this general behaviour is seen in some distributions relevant to the 
$t\bar tHH$ channel, where the NLO curves lie between the NLO+PS ones,
which can differ up to 10\% (still within the scale 
uncertainties). However, in all these cases, the corresponding
differences at the LO+PS level are systematically larger than for NLO+PS,
which thus confirms the stabilisation trend usually seen when
higher-order corrections are included.

NLO corrections in the gluon-gluon fusion channel are important rate-wise, 
yet the shapes are not strongly affected, as is apparent from the rather 
flat $K$-factors.  NLO corrections in VBF production are of order 10-20\%,
and modify the shape of the distributions towards low mass-scale values.  
NLO effects in $t\bar tHH$ production lead to a drastic reduction of the 
scale uncertainties, and to minor changes in the shapes, except for $m_{HH}$.
The associated vector boson channels display very similar features: rather
flat $K$-factors for all the distributions studied, except for $p_T(HH)$
where the NLO corrections become more and more important at high $p_T$.  

\section{Conclusions}
Assessing the nature of the newly discovered boson will need a campaign of
measurements to be performed at the LHC at an unprecedented accuracy. One of
the key processes in this endeavour is Higgs-pair production. Not only it
gives one the possibility of measuring the value of the Higgs self-coupling
$\lambda$, but also of putting constraints on several, still viable,
new-physics scenarios.  All such measurements will need accurate SM
predictions for total cross sections (in order to extract information on the
couplings) and differential distributions (in order to establish acceptances
and identify optimal selection cuts), including reliable estimates of the
theoretical uncertainties.

In this Letter we have presented the first predictions at the NLO accuracy
matched with parton shower for all the relevant Higgs-pair production channels
in the Standard Model. We find that, as expected, including NLO corrections
leads to a reduction of the theoretical uncertainties, especially significant
in the gluon-initiated channels, and provides one with reliable predictions
for the kinematic distributions of final state particles.  With the
exception of the gluon-gluon fusion process, which needs an ad-hoc treatment 
and for which a dedicated procedure and code have been developed~\cite{Vetal}, 
all the results presented here can be easily and automatically reproduced 
with the publicly available version of the {\sc MadGraph5\Q{_}aMC@NLO}
code~\cite{MG5_aMC}.

The extension of our study to models that feature physics beyond the SM is in
progress.

\section*{Acknowledgements}
\noindent
This work has been supported in part by the ERC grant 291377 "LHC Theory", by
the Swiss National Science Foundation (SNF) under contracts 200020-138206
and 200020-149517, and
grant PBELP2\_146525, by the Research Executive Agency (REA) of the European
Union under the Grant Agreement numbers PITN-GA-2010-264564 (LHCPhenoNet) and
PITN-GA-2012-315877 (MCNet). The work of FM and OM is supported by the IISN
``MadGraph'' convention 4.4511.10, by the IISN ``Fundamental interactions''
convention 4.4517.08, and in part by the Belgian Federal Science Policy Office
through the Interuniversity Attraction Pole P7/37. OM is 
"Chercheur scientifique logistique postdoctoral F.R.S.-FNRS".

\bibliographystyle{apsrev4-1}
\bibliography{library}
\end{document}